\documentclass[%
aps, 
pre,
amsmath,amssymb,
reprint,%
longbibliography
]{revtex4-1}
\usepackage{graphicx}
\usepackage{dcolumn}
\usepackage{bm}

\usepackage{amsmath}	
\begin{document}

\newcommand{\diff}[2]{\frac{d#1}{d#2}}
\newcommand{\pdiff}[2]{\frac{\partial #1}{\partial #2}}
\newcommand{\fdiff}[2]{\frac{\delta #1}{\delta #2}}
\newcommand{\bx}{\bm{x}}
\newcommand{\bq}{\bm{q}}
\newcommand{\br}{\bm{r}}
\newcommand{\bu}{\bm{u}}
\newcommand{\by}{\bm{y}}
\newcommand{\bh}{\bm{h}}
\newcommand{\bxi}{\bm{\xi}}
\newcommand{\bY}{\bm{Y}}
\newcommand{\bF}{\bm{F}}
\newcommand{\new}{\nonumber\\}
\newcommand{\abs}[1]{\left|#1\right|}
\newcommand{\tr}{{\rm Tr}}
\newcommand{\HH}{{\mathcal H}}
\newcommand{\II}{{\mathcal I}}
\newcommand{\WW}{{\mathcal W}}
\newcommand{\OO}{{\mathcal O}}
\newcommand{\FF}{{\mathcal F}}
\newcommand{\ave}[1]{\left\langle #1 \right\rangle}
\newcommand{\im}{{\rm Im}}
\newcommand{\re}{{\rm Re}}
\newcommand{\ke}{k_{\rm eff}}
\newcommand{\ipr}{{\rm IPR}}
\newcommand{\TT}{\mathcal{T}}
\newcommand{\tg}{\tilde{\Gamma}}
\newcommand{\tphi}{\tilde{\phi}}

\preprint{AIP/123-QED} \title{Continuous symmetry breaking of
low-dimensional systems driven by inhomogeneous oscillatory driving
forces
%
}


\author{Harukuni Ikeda}
 \email{harukuni.ikeda@gakushuin.ac.jp}
\affiliation{ Department of Physics, Gakushuin University, 1-5-1 Mejiro,
Toshima-ku, Tokyo 171-8588, Japan}

\author{Yuta Kuroda}
\affiliation{Department of Physics, Nagoya University, Nagoya 464-8602, Japan}

\date{\today}

\begin{abstract}
The driving forces of chiral active particles and deformations of cells
are often modeled by spatially inhomogeneous but temporally periodic
driving forces. Such inhomogeneous oscillatory driving forces have only
recently been proposed in the context of active matter, and their
effects on the systems are not yet fully understood.  In this work, we
theoretically study the impact of spatially inhomogeneous oscillatory
driving forces on continuous symmetry breaking.  We first analyze the
linear model for the soft modes in the ordered phase to derive the lower
critical dimension of the model, and then analyze the spherical model to
investigate more detailed phase behaviors.  Interestingly, our analysis
reveals that symmetry breaking occurs even in one and two dimensions,
where the Hohenberg--Mermin--Wagner theorem prohibits continuous
symmetry breaking in equilibrium. Furthermore, fluctuations of conserved
quantities, such as density, are anomalously suppressed in the
long-wavelength, {\it i.e.}, show hyperuniformity.
\end{abstract}

\maketitle

\section{Introduction}
As the temperature is lowered, gas becomes liquid and the liquid becomes
solid. These dramatic changes in physical properties are called phase
transitions~\cite{nishimori2010elements}. The phase transitions in
equilibrium systems have been actively studied using powerful tools of
statistical mechanics such as the renormalization
groups~\cite{wilson1971}, scaling theories~\cite{kadanoff1966}, exact
solutions~\cite{onsager1944,history1967}, and extensive numerical
simulations~\cite{alder1957phase,hukushima1996exchange,frenkel2001understanding}.
The phase transitions in non-equilibrium systems are also actively
investigated, though the theoretical development is still in its infancy
compared to that of the equilibrium phase transitions.

One of the most well-known no-go theorems in the theory of equilibrium
phase transitions is the Hohenberg--Mermin--Wagner
theorem~\cite{mermin1966,hohenberg1967}.  This theorem claims that
short-range interacting systems having continuous degrees of freedom do
not show long-range order in one and two dimensions.  On the other hand,
the theorem does not hold for non-equilibrium systems, and indeed there
are several examples of low-dimensional systems showing the long-range
order. Examples include the XY model driven by anisotropic
noise~\cite{xy1995,reichl2010}, $O(n)$ model driven by
shear~\cite{corberi2002,nakano2021,ikeda2401scaling}, and models driven
by anti-correlated noise~\cite{leonardo2023,ikeda2023cor}, polar active
fluids~\cite{vicsek1995,toner1995,ikeda2024minimum}, and non-reciprocal
systems~\cite{loos2022long,dadhichi2020}.

A popular class of non-equilibrium systems is periodically driven
systems.  For instance, spin systems in an AC field have been
investigated extensively to understand hysteresis of
ferromagnets~\cite{rao1989hysteresis,rao1990magnetic,dhar1992hysteresis,thomas1993hysteresis}.
Other examples include tapping experiments of granular
systems~\cite{knight1995}, and emulsions subjected to an oscillatory
shear strain~\cite{hebraud1997,berthier2001}.  In the above examples,
the systems are driven by the spatially {\it homogeneous} driving
force. For instance, in a typical setting of magnetic hysteresis, all
spins $i=1,\cdots, N$ are driven by the same external field such as
$h_i(t)=h(t)=a\sin(\omega_0 t)$, where $a$ and $\omega_0$ denote the
strength and frequency of the driving force, respectively.  Recently, a
different type of driving forces have been proposed in the context of
active matter, namely, spatially {\it inhomogeneous} but temporally
periodic driving forces~\cite{berthier2019lectures}. For instance,
tissues are often driven by oscillatory deformations of constituent
cells~\cite{zehnder2015cell,PhysRevE.96.050601}. To model this behavior,
Tjhung and Berthier introduced a model consisting of actively-deforming
particles~\cite{PhysRevE.96.050601}. The diameter $\sigma_i$ of the
$i$-th particle of the model oscillates around its mean value
$\sigma_i^0$ as follows:
\begin{align}
\sigma_i(t)=\sigma_i^0(1+ a \cos(\omega_0t + \psi_i)),\label{132331_3Nov23}
\end{align}
where $a$ and $\omega_0$ denote the strength and frequency of the
oscillation, respectively, and $\psi_i$ denotes a random phase
shift~\cite{PhysRevE.96.050601}. When $\psi_i=0$,
Eq.~(\ref{132331_3Nov23}) represents the affine
compression/decompression, while when $\psi\neq 0$, it leads spatially
inhomogeneous deformations. Another example of inhomogeneous driving
force appears in chiral active matter, where constituent particles
spontaneously rotate due to the periodic nature of the driving
force~\cite{chiral2008,debets2023}. A popular numerical model to
describe the rotational motion of a chiral active particle
in two dimensions 
is~\cite{Callegari2019}
\begin{align}
&\dot{x}_i = -\pdiff{V}{x_i}+h\cos\theta_i +\xi_{i,x},\new
&\dot{y}_i = -\pdiff{V}{y_i}+h\sin\theta_i + \xi_{i,y},\new
&\dot{\theta}_i = \omega_0 + \xi_\theta,
\end{align}
where $(x_i,y_i)$ denotes position of the $i$-th particle,
$\xi_{i,x,y,\theta}$ denotes the noise, and $V$ denotes the interaction
potential. In the absence of the noise $\xi_{i,x,y,\theta}=0$, the
equations are reduced to
\begin{align}
&\dot{x}_i = -\pdiff{V}{x_i}+h_{i,x},
&\dot{y}_i = -\pdiff{V}{y_i}+h_{i,y}.
\end{align}
Here the particle is driven by the inhomogeneous driving force
$h_{i,x}=h\cos(\omega_0 t + \psi_i)$ and $h_{i,y}=h\sin(\omega_0 t +
\psi_i)$, where the phase shift $\psi_i$ is determined by the initial
condition $\psi_i=-\omega_0 t_{\rm ini}+\theta_i(t_{\rm ini})$.  The
phase behavior of this type of system, including actively deforming
particles and chiral active matter, has been studied
recently~\cite{liebchen2022chiral,zhi2020,huang2021circular,
Lin2022,semwal2022macro, PhysRevE.96.050601,zhang2023, li2024}.
However, the effects of the inhomogeneous driving force on the phase
transition have not been fully understood yet.

This study focuses on continuous symmetry breakings such as
crystallization and nematic transition systems driven by inhomogeneous
oscillatory driving forces.  In particular, we show that continuous
symmetry-breaking can occur even in low dimensions $d\leq 2$ for systems
driven solely by inhomogeneous oscillatory driving forces.  To achieve
this goal, we first investigate the linear model and derive the lower
critical dimension.  Here we only assume that the model has at least one
massless mode in the ordered phase, and thus, the result should be
applied for general types of continuous symmetry breaking such as
crystallization and nematic phase transitions.  Next, we investigate the
spherical model in order to discuss more detailed phase
behaviors~\cite{spherical1966,nishimori2010elements}.

The spherical model was first introduced by Berlin and Kac as a
simplification of the Ising model~\cite{berlin1952spherical}.  The model
can be solved analytically in any dimension, both in equilibrium and
nonequilibrium~\cite{stanley1968,nishimori2010elements,henkel2008non}.
In equilibrium, the model undergoes the second-order phase transition
from the paramagnetic to the ferromagnetic phase at finite
temperatures. The critical exponents of the transition agree with those
of the $n\to\infty$ limit of the $O(n)$
model~\cite{nishimori2010elements}.  In previous studies, the spherical
model driven by the homogeneous oscillatory driving forces has been
investigated extensively in the context of the magnetic
hysteresis~\cite{rao1989hysteresis,rao1990magnetic,dhar1992hysteresis,thomas1993hysteresis},
and also structural glasses~\cite{berthier2001}. However, to the best of
our knowledge, the spherical model driven by the inhomogeneous
oscillatory driving force has not been investigated so far.

The structure of the manuscript is as follows.  In
Sec.~\ref{114008_4Nov23}, we investigate the linear model by using the
scaling analysis. In Sec.~\ref{114029_4Nov23}, we investigate the
spherical model. In Sec.~\ref{083910_21Apr23}, we shortly discuss the
behavior of the conserved order parameter such as density. In
particular, we show that the fluctuations of the conserved quantity are
highly suppressed, implying that the model exhibits
hyperuniformity~\cite{torquato2018hyperuniform}. In
Sec.~\ref{114401_4Nov23}, we conclude the work.

\section{Linear model}
\label{114008_4Nov23}

When continuous symmetries are spontaneously broken, there arise soft
modes called the Nambu--Goldstone (NG)
modes~\cite{nambu1960,goldstone1962}. In equilibrium, the fluctuations
of NG modes diverge in $d\leq 2$, which destroys the long-range order.
Therefore, continuous symmetries are not spontaneously broken for $d\leq
2$ in equilibrium systems.  This fact is nowadays widely known as the
Hohenberg--Mermin--Wagner theorem~\cite{hohenberg1967,mermin1966}. We
here investigate the scaling behavior of the NG modes to discuss the
stability of the ordered phase of systems driven by inhomogeneous
oscillatory driving forces. Let $u$ be the displacement along one of the
NG modes. We assume that $u$ follows the following phenomenological
linear equation:
\begin{align}
&\pdiff{u(\bx,t)}{t} = \nabla^2u(\bx,t) + f(\bx,t),\label{185404_5Aug23}\\
&f(\bx,t)=\xi(\bx,t)+ h(\bx,t)\label{drive},
\end{align}
where $\xi(\bx,t)$ denotes a Gaussian white noise satisfying
\begin{align}
&\ave{\xi(\bx,t)} = 0,\new
&\ave{\xi(\bx,t)\xi(\bx',t')} = 2T\delta(\bx-\bx')\delta(t-t'), \label{noise}
\end{align}
and $h(\bx,t)$ represents an inhomogeneous oscillatory driving force:
\begin{align}
h(\bx,t) = \sqrt{2D}\cos(\omega_0t+\psi(\bx))\abs{d\bx}^{-1/2},\label{161706_18Nov23}
\end{align}
where $\abs{d\bx}$ denotes the volume of the unit cell of the
discretization.  $T$ and $D$ stand for the strength of the noise and the
driving force, respectively.  In Eq.~(\ref{161706_18Nov23}), we
introduced the random phase shift $\psi(\bx)$ distributed uniformly in
$[0,2\pi]$~\cite{berthier2019lectures}.  The case with $\psi(\bx)=0$
corresponds to a homogeneous driving force. The mean and variance of
$h(\bx,t)$ are~\cite{zwanzig2001nonequilibrium}
\begin{align}
&\ave{h(\bx,t)} =0, \new 
&\ave{h(\bx,t)h(\bx',t')}= D\delta(\bx-\bx')\cos(\omega_0(t-t')),
\label{112528_24Apr23}
\end{align}
where we used $\delta(\bx-\bx')=\lim_{\abs{d\bx}\to
0}\delta_{\bx,\bx'}/\abs{d\bx}$ in the continuum limit.  In
Eqs.~(\ref{noise}) and (\ref{112528_24Apr23}), the symbol
$\ave{\bullet}$ represents either the average over the noise $\xi$ or
the randomness $\psi$.  Equation~(\ref{112528_24Apr23}) implies that the
time translational symmetry is restored after taking the average for the
quenched randomness $\psi(\bx)$.  To investigate the scaling behavior,
we consider the following scaling
transformation~\cite{burger1989,toner1995}
\begin{align}
x\to bx,\ t\to b^{z}t,\ u\to
b^{\chi}u.
\end{align}
To calculate the scaling dimension of the driving force $f(\bx,t)$, we
observe the fluctuation induced by $f(\bx,t)$ in $d+1$ dimensional
Euclidean space $[0,l]^d\times [0,t]$~\cite{torquato2018hyperuniform}:
\begin{align}
\sigma(l,t)&=
\ave{\left(\int_{\bx'\in [0,l]^d}
 \hspace{-5mm}d\bx' \int_0^t dt'f(\bx',t')\right)^2}\new 
&= \sigma_\xi(l,t) + \sigma_h(l,t),
\end{align}
where 
\begin{align}
\sigma_\xi(l,t)&=
\ave{\left(\int_{\bx'\in [0,l]^d}
 \hspace{-5mm}d\bx' \int_0^t dt'\xi(\bx',t')\right)^2},\new
\sigma_h(l,t)&=
\ave{\left(\int_{\bx'\in [0,l]^d}
 \hspace{-5mm}d\bx' \int_0^t dt'h(\bx',t')\right)^2}.
\end{align}
For $T>0$ and $\omega_0\neq 0$, we get $\sigma_\xi \sim l^{d}t$ and
$\sigma_h\sim l^{d}t^0$, leading to $\sigma \sim \sigma_\xi \sim
l^{d}t$ for $l\gg 1$ and $t \gg 1$. Since $\sigma \sim t^2 l^{2d}f^2$,
we have ${f(\bx,t)\to b^{-z/2}b^{-d/2}f(\bx,t)}$ for $T>0$ and
$\omega_0\neq 0$. On the contrary, for $T=0$ and $\omega_0\neq 0$, we
get $\sigma=\sigma_h \sim l^{d}t^0$, leading to ${f(\bx,t)\to
b^{-z}b^{-d/2}f(\bx,t)}$.  For $\omega_0=0$, $h(\bx,t)\propto
\cos\psi(\bx)$ is a time-independent constant force, which leads to the
ballistic behavior $\sigma\sim \sigma_h\sim l^{d}t^2$. Therefore, we
get ${f(\bx,t)\to b^{-d/2}f(\bx,t)}$ for $\omega_0=0$.  In summary,
after the scaling transformation, Eq.~(\ref{185404_5Aug23}) reduces to
\begin{align}
&b^{\chi-z}\pdiff{u(\bx,t)}{t} = b^{\chi-2}\nabla^2u(\bx,t) + b^{z_f}f(\bx,t),
\new 
&z_f  = 
\begin{cases}
 -\frac{z}{2}-\frac{d}{2} & T>0,\omega_0\neq 0\\
 -z -\frac{d}{2} & T=0,\omega_0\neq 0\\
 -\frac{d}{2} & \omega_0=0
\end{cases}.\label{175326_4Feb24}
\end{align}
Assuming all terms have the same scaling dimension, we get the following
scaling relations~\cite{nishimori2010elements,maggi2022critical}:
\begin{align}
 \chi-z = \chi-2 = z_f,
\end{align}
leading to 
\begin{align}
&z = 2,\new
&\chi =
\begin{cases}
1-\frac{d}{2}& T>0, \omega_0\neq 0\\
-\frac{d}{2}& T=0, \omega_0\neq 0\\
2-\frac{d}{2}& \omega_0=0
\end{cases}.\label{213312_7Nov23}
\end{align}
To see the stability of the ordered phase, 
we observe the fluctuations of the NG modes~\cite{toner1995}:
\begin{align}
\ave{\delta u^2}\sim b^{2\chi}.
\end{align} 
For the ordered phase to be stable, $\chi$ must be negative; otherwise,
the fluctuations diverge in the thermodynamic limit $b\to\infty$, and
the long-range order is destroyed~\cite{toner1995}. Therefore, the
lower-critical dimension can be determined by setting $\chi=0$:
\begin{align}
d_l =
\begin{cases}
2& T>0, \omega_0\neq 0\\
0& T=0, \omega_0\neq 0\\
4& \omega_0=0
\end{cases}.\label{113022_6Aug23}
\end{align}
The above equation is the main result of this paper.  For $T>0$ and
$\omega_0\neq 0$, we get $d_l=2$, as in
equilibrium~\cite{mermin1966,hohenberg1967}, meaning that the periodic
driving forces do not change the lower critical dimension at finite
$T$. On the contrary, for $T=0$ and $\omega_0\neq 0$, we get $d_l=0$,
meaning that the long-range order can appear even in $d=1$ and $d=2$.
For $\omega_0=0$, we get $d_l=4$.  This is natural since
$h(\bx,t)\propto \cos\psi(\bx)$ plays the role of the random field,
which destroys the order in the systems with continuous symmetry for
$d\leq 4$ according to the Imry--Ma
arguments~\cite{imry1975,nishimori2010elements}.
Equation~(\ref{113022_6Aug23}) should also be applied for an arbitrary
function $h(\bx,t)$ that is temporally periodic and spatially
uncorrelated for which $\sigma_h\sim l^{d/2}t^0$.

The argument so far is phenomenological but quite general, where we only
assume the existence of the NG modes. Eq.~(\ref{113022_6Aug23}) should
be applied for general types of continuous symmetry breaking such as the
ferromagnetic phase transition of the $O(n)$ model with $n\geq 2$,
crystalization, and nematic phase transition.

However, the above linear model cannot describe the phase transitions.
To investigate the phase behaviors more closely, in the next section, we
investigate the mean-spherical model~\cite{berlin1952spherical}, which
corresponds to the $n\to\infty$ limit of the $O(n)$ model and can be
solved exactly~\cite{nishimori2010elements}.


\section{Spherical model}
\label{114029_4Nov23}

For concreteness, we here introduce and investigate the mean-spherical
model and discuss more detailed phase behaviors. Let $\phi(\bx,t)$ be a
non-conserved order parameter such as the magnetization. We assume that
the time evolution of $\phi(\bx,t)$ follows the model-A
dynamics~\cite{hohenberg1977}:
\begin{align}
&\pdiff{\phi(\bx,t)}{t} = -\fdiff{\FF[\phi]}{\phi(\bx,t)}
+f(\bx,t),
 \label{141313_15Feb23}
\end{align}
where the driving force $f(\bx,t)$ is defined by
Eq.~(\ref{drive}). The free-energy $\FF[\phi]$ 
of the mean-spherical model 
is defined as~\cite{crisanti2019,ikeda2023cor}:
\begin{align}
\FF[\phi] = \int d\bx \left[
 \frac{(\nabla\phi)^2}{2}+
 \frac{\mu\phi^2}{2}\right],\label{032725_17Mar23}
\end{align}
where $\mu$ denotes the Lagrange multiplier to impose the mean-spherical
constraint~:
\begin{align}
\int d\bx \ave{\phi(\bx)^2} = N.
\end{align}
The model has been studied extensively in equilibrium $D=0$, where the
critical exponents agree with those of the large $n$ limit of the $O(n)$
model~\cite{henkel2008non,nishimori2010elements}.  Below we investigate
the phase behaviors of the model in the thermodynamics limit:
$N\to\infty$ and $\int d\bx = V\to\infty$ with fixed ratio $\rho=N/V$.

We shall make some comments on $\mu$.
Equation~(\ref{141313_15Feb23}) is written as
\begin{align}
\pdiff{\phi(\bx,t)}{t} = -(\mu-\nabla^2)\phi(\bx,t)
+f(\bx,t).\label{213828_16Apr23}
\end{align}
By integrating Eq.~(\ref{213828_16Apr23}) w.r.t. $\bx$, we get
$\dot{m}(t) = -\mu m(t)$, where $m(t)=V^{-1}\int d\bx \phi(\bx,t)$. The
Lagrange multiplier should satisfy $\mu\geq 0$ since otherwise, the
steady state becomes unstable. In the steady state $\dot{m}=-\mu m=0$.
This condition is automatically satisfied in the disordered phase $m=0$.
In the ordered phase $m>0$, on the contrary, $\mu$ should vanish
$\mu=0$.  Therefore, in the ordered phase Eq.~(\ref{213828_16Apr23})
reduces to the linear equation Eq.~(\ref{185404_5Aug23}) analyzed in the
previous section. As a consequence, the lower critical dimension $d_l$
of the mean-spherical model is given by Eq.~(\ref{113022_6Aug23}), as we
will confirm later.

We decompose the order parameter $\phi(\bx,t)$ into the mean value $m$
and fluctuations $\tilde{\phi}(\bx,t)\equiv\phi(\bx,t)-m$. Since $\dot m
= -\mu m = 0$ in the steady state, $\tilde{\phi}(\bx,t)$ satisfies the
same equation as Eq.~(\ref{213828_16Apr23}). Thus, the Fourier transform
of the equation leads to
\begin{align}
i\omega \tphi(\bq,\omega)
= -(q^2+\mu)\tphi(\bq,\omega) + f(\bq,\omega),\label{190005_18Apr23}
\end{align}
where $q=\abs{\bq}$ and 
\begin{align}
\tphi(\bq,\omega) = \int d\bx dt e^{-i\bq\cdot\bx-i\omega t}
\tilde{\phi}(\bx,t),
\end{align}
From Eq.~(\ref{190005_18Apr23}), we get
\begin{align}
\tphi(\bq,\omega) = \frac{f(\bq,\omega)}{i\omega + (q^2+\mu)}.
\end{align}
The two-point correlation is then calculated as
\begin{align}
\ave{\phi(\bq,\omega)\phi(\bq',\omega')}
&= (2\pi)^{d+1}\delta(\bq+\bq')\delta(\omega+\omega')S(\bq,\omega),
\end{align} 
where $\phi(\bq,\omega)$ denotes the Fourier transform of $\phi(\bx,t)$,
and
\begin{align}
S(\bq,\omega) 
&=m^2 (2\pi)^{d+1}\delta(\omega)\delta(\bq)
 + \frac{2T}{\omega^2+(q^2+\mu)^2}\new 
 &+\frac{D\pi}{\omega_0^2+(q^2+\mu)^2}
 \left[\delta(\omega+\omega_0)+\delta(\omega-\omega_0)\right]. 
\label{Sqomega_modelA}
\end{align}
The static correlation is obtained by integrating
Eq.~(\ref{Sqomega_modelA}) over $\omega$:
\begin{align}
S(\bq) &=
\frac{1}{2\pi}\int d\omega S(\bq,\omega)\new 
&= (2\pi)^d m^2\delta(\bq)+ \frac{T}{q^2+\mu}
+ \frac{D}{\omega_0^2+(q^2+\mu)^2}. \label{Sq_modelA}
\end{align}

The remaining task is to determine the Lagrange multiplier $\mu$ by the
spherical constraint:
\begin{align}
N=\int d\bx \ave{\phi(\bx,t)^2}
  = \frac{V}{(2\pi)^d}\int d\bq S(\bq).\label{042636_17Mar23}
\end{align}
Substituting Eq.~(\ref{Sq_modelA}) into Eq.~(\ref{042636_17Mar23}), we
get
\begin{align}
1 = \frac{m^2}{\rho} +
TF(\mu)+ DG(\mu),\label{223510_16Apr23}
\end{align}
where 
\begin{align}
F(\mu) 
 &=\frac{\Omega_d}{(2\pi)^d\rho}
\int_0^{q_D}\frac{dqq^{d-1}}{q^2+\mu},\new
G(\mu)
&= \frac{\Omega_d}{(2\pi)^d\rho}\int_0^{q_D}\frac{dqq^{d-1}}{\omega_0^2+(q^2+\mu)^2}.
\label{100701_5Aug23}
\end{align}
Here $q_D$ denotes the phenomenological cutoff, and $\Omega_d$
denotes the $d$-dimensional solid angle.

\begin{figure}[t]
\includegraphics[width=9cm]{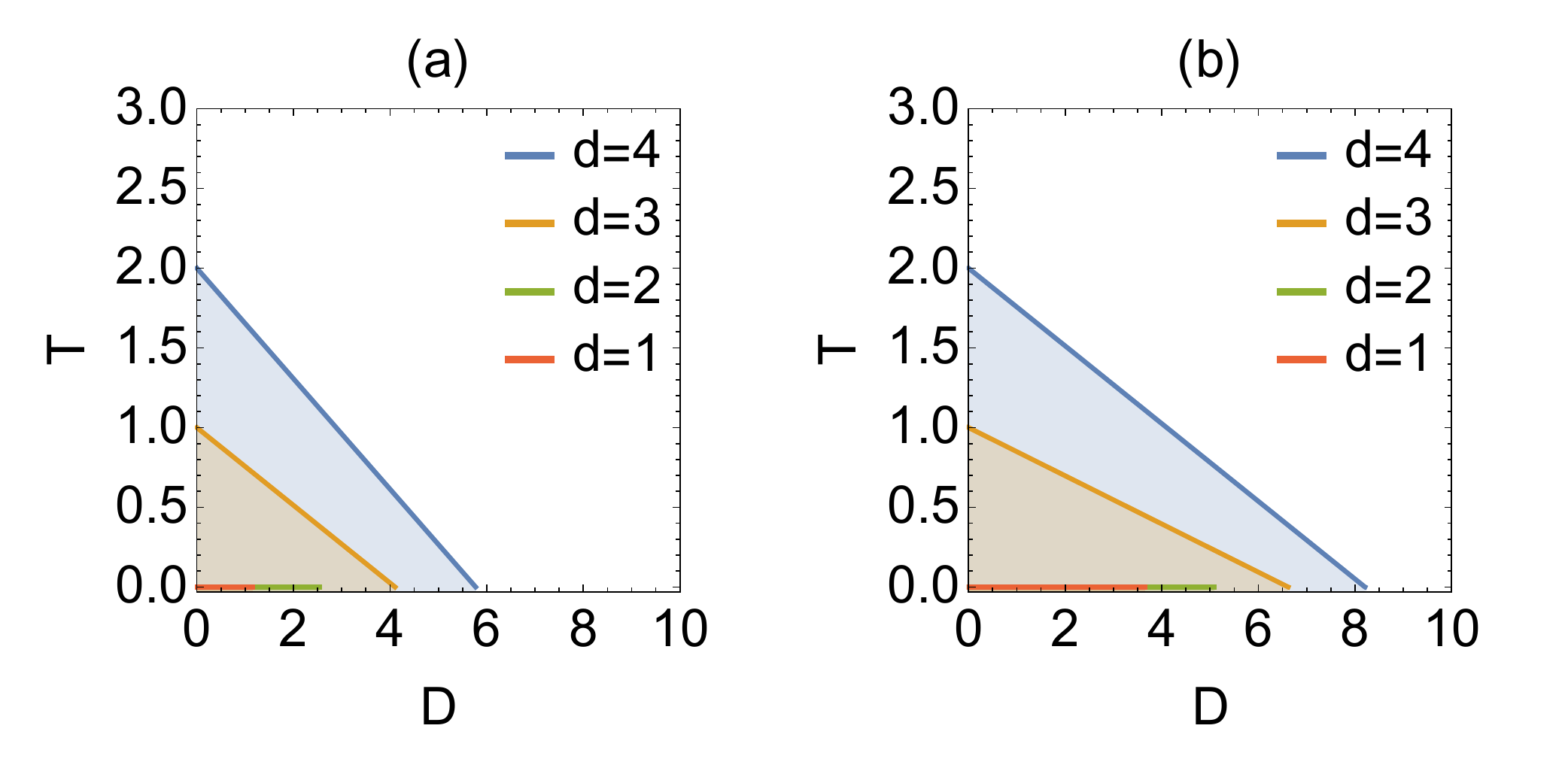} \caption{ $T$--$D$ phase diagrams
for model-A (a) and model-B (b). The solid lines denote the critical
lines $T_c$ for $d=4$, $3$, $2$, and $1$ from top to bottom.
The filled regions
represent the ordered phase.  For $d\leq 2$, ordered phases appear only
at $T=0$. For simplicity here we set $\rho=\Omega_d/(2\pi)^d$, $q_D=1$,
and $\omega_0=1$.}  \label{103304_19Apr23}
\begin{center}
\end{center}
\end{figure}

Now we discuss the phase behavior for finite $T$.  For sufficiently high
$T$, the system is in a disordered phase.  As $T$ decreases, the model
undergoes a continuous phase transition at a certain point $T=T_c$.
Since $\mu=0$ and $m=0$ at the transition point, $T_c$ is calculated as
\begin{align}
T_c = \frac{1-DG(0)}{F(0)}.\label{110448_19Apr23}
\end{align}
For $d\leq 2$, $F(0)$ diverges, and thus the phase transition does not
occur at finite $T$. In other words, the lower critical dimension is
$d_l=2$. This is the same situation as that in the equilibrium phase
transition, where the Hohenber-Mermin-Wagner theorem prohibits the
ordered phase for $d\leq 2$~\cite{mermin1966,hohenberg1967}.  We show
$T_c$ in Fig.~\ref{103304_19Apr23}~(a).

\begin{figure}[t]
\includegraphics[width=9cm]{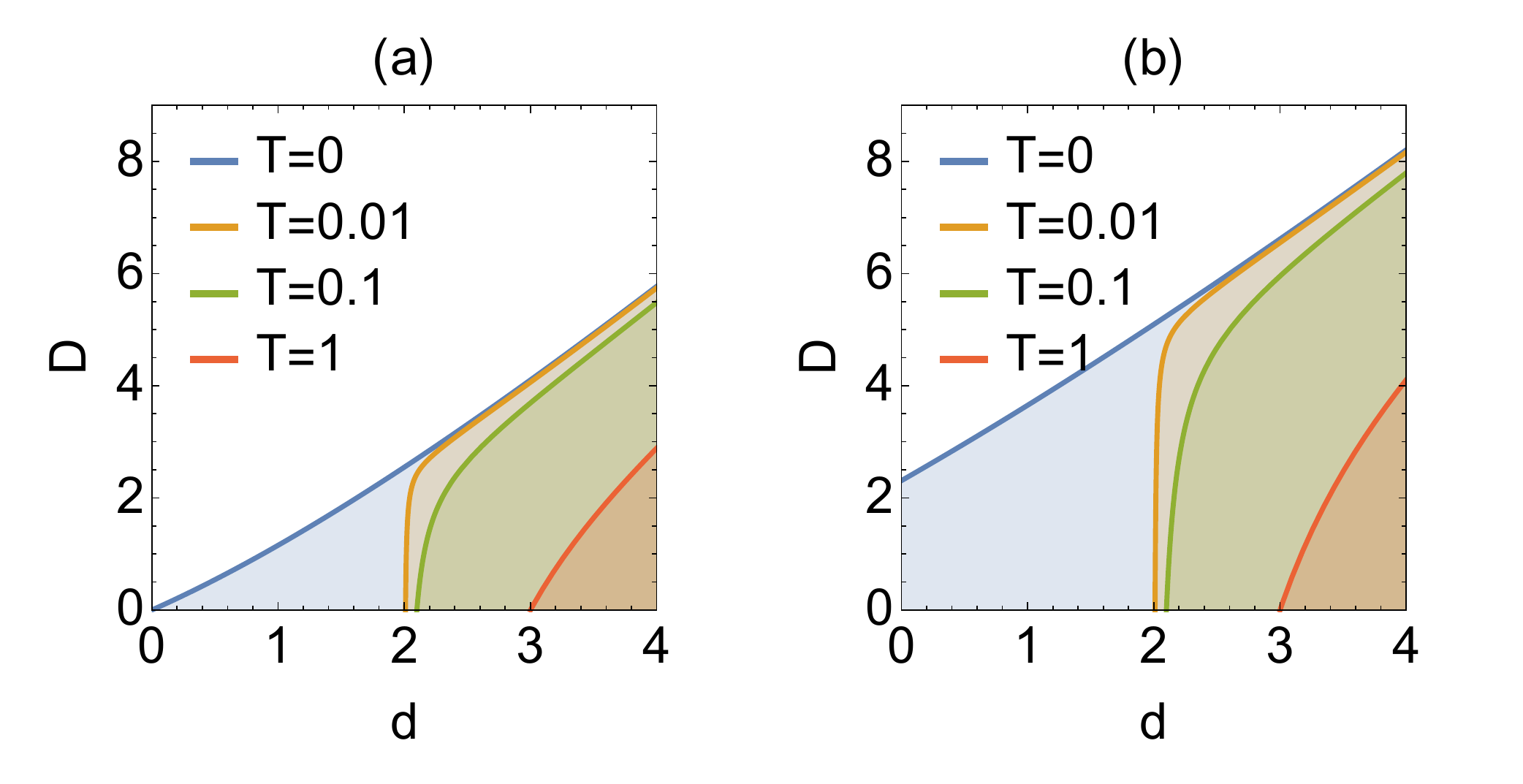} \caption{$D$--$d$ phase diagrams
for model-A (a) and model-B (b). The solid lines represent $D_c$ for
$T=0$, $0.01$, $0.1$, and $1$ from top to bottom. The filled regions represent
the ordered phases. When $T>0$, the ordered phase appears only if
$d>2$. On the contrary, when $T=0$, the ordered phase appears for all
$d>0$. For simplicity we here set $\rho=\Omega_d/(2\pi)^d$, $q_D=1$, and
$\omega_0=1$.}  \label{142158_18Apr23}
\end{figure}

Next, we discuss the phase behavior in the limit $T\to
0$. For this purpose, it is convenient to control $D$ instead of $T$.
For sufficiently large $D$, the system is in the disordered phase.
With decreasing $D$, the model undergoes the phase transition at
\begin{align}
 D_c = \frac{1-TF(0)}{G(0)}.
\end{align}
For $T>0$ and $d\leq 2$, $F(0)$ diverges, and thus the ordered phase
does not appear. On the contrary, when $T=0$, $D_c$ can have a finite
value $\lim_{T\to 0}D_c=1/G(0)$ for all $d>0$, which implies $d_l=0$ and
is consistent with Eq.~(\ref{113022_6Aug23}). To visualize this result,
in Fig.~\ref{142158_18Apr23}(a), we show the dimensional dependence of
$D_c$ for various temperatures.

\begin{figure}[t]
\includegraphics[width=9cm]{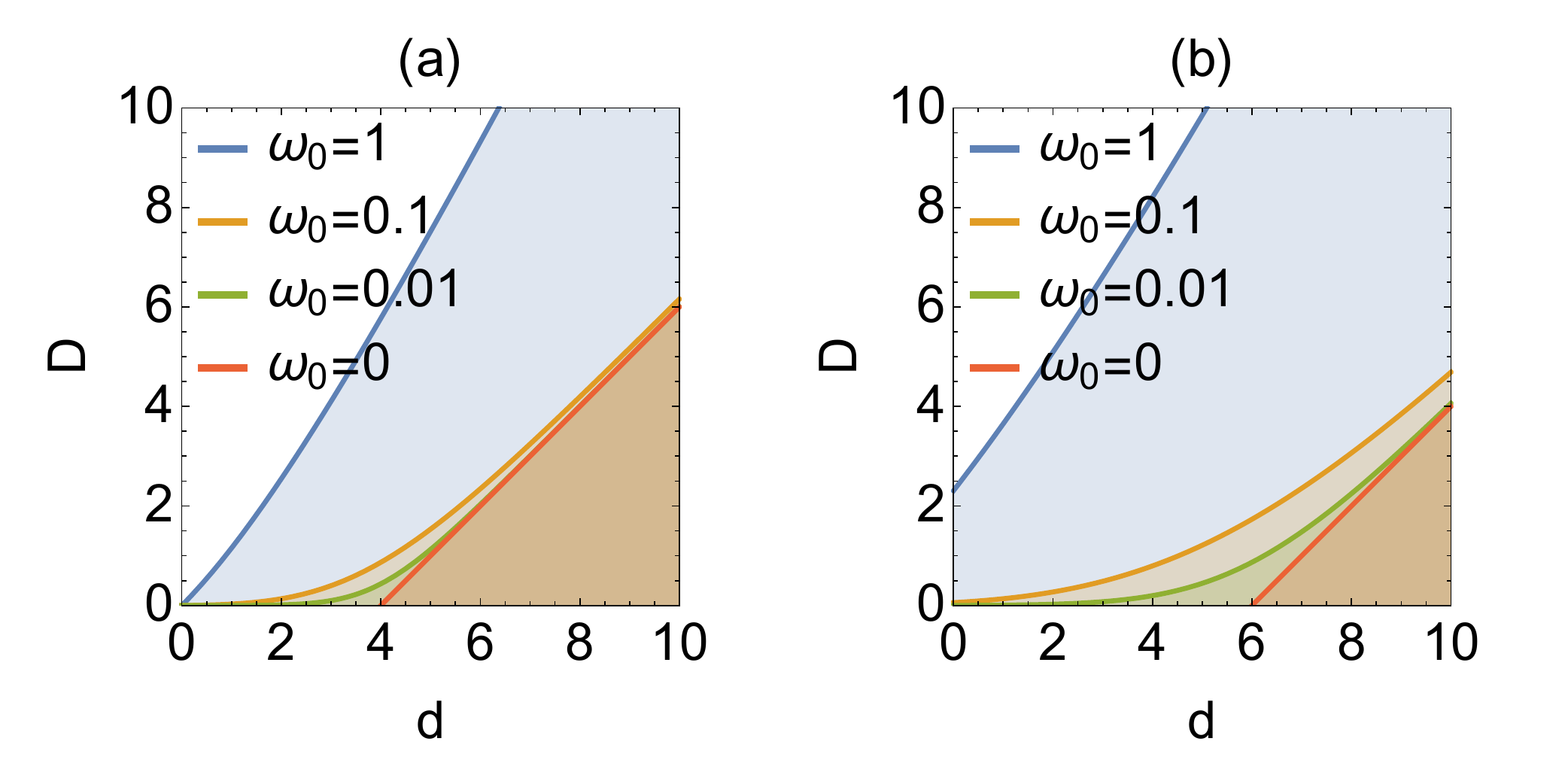} \caption{ $D$--$d$ phase diagrams
for $T=0$ for model-A (a) and model-B (b). The solid lines denote $D_c$
for $\omega_0=1$, $0.1$, $0.01$, and $0$ from top to bottom. The filled
regions represent the ordered phases. When $\omega_0>0$, the ordered
phase appears for all $d>0$. On the contrary, when $\omega_0=0$, the
ordered phase appears only when $d>4$ for the model-A and $d>6$ for the
model-B. For simplicity we here set $\rho=\Omega_d/(2\pi)^d$, $q_D=1$.}
\label{124953_21Apr23}
\end{figure}

Finally, we consider the $\omega_0$ dependence of the phase behavior. For
$\omega_0\neq 0$, all qualitative phase behaviors discussed above remain
unchanged. On the contrary, when $\omega_0=0$, $G(0)$ diverges for
$d\leq 4$, leading to $D_c\to 0$ and $T_c\to -\infty$. Therefore, the
phase transition does not occur for all $d\leq 4$.  This is a natural
result because $h(\bx,t)$ for $\omega_0=0$ plays the role of a random
field, which destroys the long-range order for $d\leq 4$ according to
the Imry--Ma argument for continuous variables~\cite{imry1975}.  In
Fig.~\ref{124953_21Apr23}~(a), we show $D$--$d$ phase diagram for
various $\omega_0$ to visualize how the ordered phase for $d\leq 4$
disappears in the limit $\omega_0\to 0$. The above result implies
$d_l=4$, which is consistent with the linear model
Eq.~(\ref{113022_6Aug23}).

\section{Conserved order parameter}

\label{083910_21Apr23}

\begin{figure}[t]
\includegraphics[width=7cm]{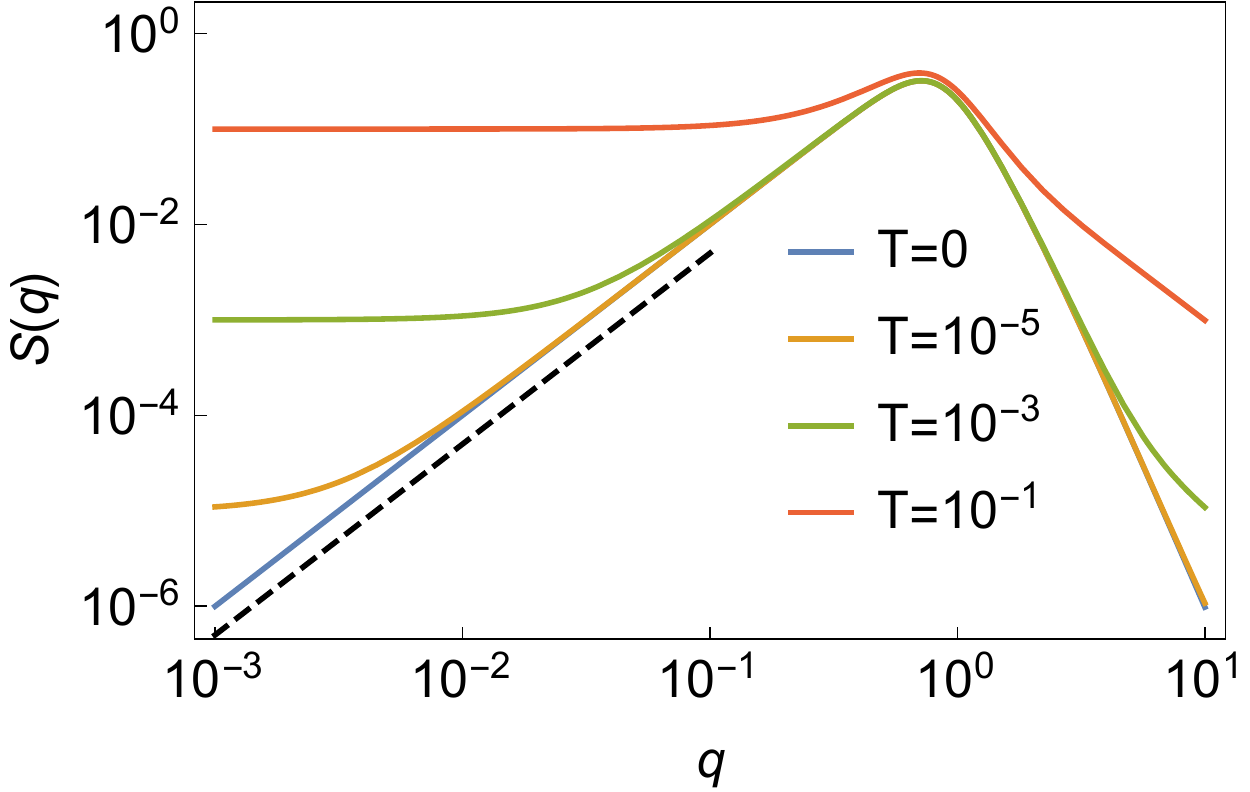} \caption{$S(q)$ for model-B. The
solid lines denote $S(q)$ for $T=0$, $10^{-5}$, $10^{-3}$, $10^{-1}$
from bottom to top. The black dashed line denotes $S(q)\propto
q^2$. When $T=0$, the model exhibits hyperuniformity $S(q)\propto q^2$
for $q\ll 1$. For simplicity, here we set $D=1$, $\omega_0=1$, and
$\mu=1$. } \label{133922_19Apr23}
\begin{center}
\end{center}
\end{figure}
So far, we discussed the phase behavior for the non-conserved order
parameter.  Below we discuss what will happen for the conserved order
parameter such as density.  A conserved quantity $\phi(\bx,t)$ follows
the continuity equation:
\begin{align}
 \pdiff{\phi(\bx,t)}{t} = -\nabla\cdot\bm{J}(\bx,t),
\end{align}
where $\bm{J}$ denotes the flux. In the model-B dynamics, $\bm{J}$ is
written as~\cite{hohenberg1977,onuki2002phase,cates2019active}
\begin{align}
&\bm{J}(\bx,t) = -\nabla\fdiff{\FF[\phi]}{\phi(\bx,t)} + \bm{f}(\bx,t), \new 
&\bm{f}(\bx,t) = \bm{\xi}(\bx,t) + \bm{h}(\bx,t),
\end{align}
where $\FF[\phi]$ is the free-energy of the mean-spherical model Eq.~(\ref{032725_17Mar23}),
and $\bm{\xi}=\{\xi_a\}_{a=1,\cdots,d}$
and $\bm{h}=\{h_a\}_{a=1,\cdots, d}$ satisfy 
\begin{align}
&\ave{\xi_a(\bx,t)} = 0,\new 
&\ave{\xi_a(\bx,t)\xi_b(\bx',t')} = 2T\delta_{ab}\delta(\bx-\bx')\delta(t-t'),\new 
& \ave{h_a(\bx,t)} = 0,\new 
& \ave{h_a(\bx,t)h_b(\bx',t')} = \delta_{ab}\delta(\bx-\bx')D\cos(\omega_0(t-t')).
\end{align}
A calculation very similar to that in the previous section yields
\begin{align}
 S(\bq) &=
 (2\pi)^d m^2\delta(\bq)+\frac{T}{q^2+\mu}
+\frac{Dq^2}{\omega_0^2 + q^4(q^2+\mu)^2}.
\end{align}
When $T=0$ for $q\neq 0$, we get 
\begin{align}
\lim_{T\to 0}S(\bq)=\frac{Dq^2}{q^4(q^2+\mu)^2+\omega_0^2}.\label{161622_21Apr23}
\end{align}
For $q\ll 1$, we observe $S(q)\sim q^2$: the large-scale fluctuations
are highly suppressed in the limit $T\to 0$, see
Fig.~\ref{133922_19Apr23}. This anomalous suppression of the
fluctuations is referred to as
hyperuniformity~\cite{torquato2018hyperuniform}.  Hyperuniformity has
been also reported in other periodically driven systems, such as chiral
active matter~\cite{huang2021circular,hyperchiral2022} and pulsating
epithelial tissues~\cite{li2024}. In previous theoretical works, the
authors in Refs.~\cite{lei2019hydrodynamics,lei2019nonequilibrium}
discussed that hyperuniformity of the chiral active matter is a
consequence of the conservation of the center of
mass~\cite{hexner2017noise}.  Our model instead assets that this
phenomenon is induced by the periodic nature of the driving force. A
recent theoretical calculation by the fluctuating hydrodynamics also
supports this conclusion, see
Ref.~\cite{kuroda2023microscopic}. However, it is not necessary that all
instances of hyperuniformity come from a single underlying cause. For
instance, the long-range hydrodynamic interaction is another source of
hyperuniformity~\cite{torquato2021s,huang2021circular,
oppenheimer2022hyperuniformity}. Further studies would be beneficial to
elucidate this point.

When $\omega_0=0$, $S(\bq)$ shows the power-low divergence $S(q)\sim
q^{-2}$ for $q\ll 1$, meaning that the large-scale fluctuations 
are significantly enhanced even in the disordered phase. The same power-low
behavior has been previously reported for the scalar active matter in
quenched random potentials~\cite{sunghan2021}.

The Lagrange multiplier $\mu$ is to be determined by 
\begin{align}
1 = \frac{m^2}{\rho} +
TF_B(\mu)+ DG_B(\mu),
\end{align}
where $\rho=N/V$ and 
\begin{align}
F_B(\mu) 
 &=\frac{\Omega_d}{(2\pi)^d\rho}
\int_0^{q_D}\frac{dqq^{d-1}}{q^2+\mu},\new
G_B(\mu)
&= \frac{\Omega_d}{(2\pi)^d\rho}\int_0^{q_D}\frac{dqq^{d-1}q^2}{\omega_0^2+q^4(q^2+\mu)^2}.
\label{100701_5Aug23}
\end{align}
Then, one can draw the critical line by $1=m^2/\rho + TF_B(0)+DG_B(0)$
as in the case of the model-A. Note that $m$ should be considered as the
control parameter since the model-B dynamics conserves $m$.  We here
only discuss the behavior for $m=0$. We show the phase diagrams in
Figs~1--3~(b), and the lower critical dimension is
\begin{align}
d_l =
\begin{cases}
2& T>0, \omega_0\neq 0\\
0& T=0, \omega_0\neq 0\\
6& \omega_0=0
\end{cases}.
\end{align}
For $\omega_0\neq 0$, the qualitative phase behaviors agree with that
of the model-A.  However, for $\omega_0=0$, $G_B(0)$ diverges for $d\leq
6$, leading to $d_l=6$.  Therefore, the ordered phase does not appear
for $d\leq 6$, see Fig.~\ref{124953_21Apr23}~(b). This result also
consistent with the Imry--Ma argument for the continuous symmetry
breaking of the active matter in quenched random
fields~\cite{sunghan2021}.

\section{Summary and discussions}
\label{114401_4Nov23} In summary, we have introduced and investigated
the linear and spherical model driven by the temporally periodic but
spatially inhomogeneous driving forces. We found that in the absence of
the thermal noise, the lower critical dimension is $d_l=0$, meaning that
the models show the continuous symmetry breaking even in $d=1$ and
$2$. Furthermore, the model for the conserved order parameter (model-B)
shows hyperuniformity.

It is well-known from the Imry--Ma argument that time-independent
inhomogeneous external fields prohibit the continuous symmetry breaking
for $d\leq 4$~\cite{imry1975}. One may expect that the time dependence
of the external field introduces extra complexity, which makes the
system more disordered. However, our result demonstrated that the
inhomogeneous oscillatory external field allows the continuous symmetry
breaking for $d\leq 4$. To the best of our knowledge, this
counter-intuitive result has not been reported before.

It is also surprising that our model undergoes continuous symmetry
breaking even in $d=1$.  Although there are several known examples of
non-equilibrium phase transitions in one
dimension~\cite{czirok1999,oloan1999,solon2013,evans2005nonequilibrium,henkel2008non,hexner2017noise},
the continuous symmetry breaking is hardly observed in $d=1$, though our
recent studies reported several other promising
candidates~\cite{ikeda2023cor,ikeda2023harmonic}.

To derive the lower critical dimension, Eq.~(\ref{113022_6Aug23}), we
only assumed the existence of the NG modes. Therefore, the result
$d_l=0$ should be generally applied for continuous symmetry breaking of
periodically driven systems. For instance, our theory can be tested in
crystal phases of chiral active
particles~\cite{Callegari2019,liebchen2022chiral} and actively-deforming
particles~\cite{PhysRevE.96.050601,parisi2023} by observing the
translational order parameters. We hope our results will be verified in
numerical simulations and experiments of those systems.

Note added: After the previous version was submitted, one of the authors
confirmed that chiral active particles can indeed exhibit long-range
crystalline order even in two dimensions~\cite{kuroda2024long}.

\acknowledgments
This project has received JSPS KAKENHI Grant Numbers 23K13031.

\appendix

\bibliography{reference}

\end{document}